\documentclass{optica-article}

\journal{opticajournal} 

\articletype{Research Article}

\usepackage{lineno}

\newcommand{\red}[1]{\textcolor{black}{#1}}

\usepackage{braket}
\newcommand{\ve}{\varepsilon}

\renewcommand{\i}{\text{i}}

\renewcommand{\Im}{\text{Im}}
\renewcommand{\Re}{\text{Re}}
\newcommand{\e}{\text{e}}

\newcommand{\s}{\text{s}}

\newcommand{\pp}{\perp}
\newcommand{\pa}{\parallel}

\begin{document}

\title{Sensing Birefringence and Diattenuation with Undetected Light}

\author{Cristofero Oglialoro\authormark{1,*} and Enno Giese\authormark{1}}

\address{\authormark{1}Technische Universit{\"a}t Darmstadt, Institut f{\"u}r Angewandte Physik, Schlossgartenstra{\ss}e 7, 64289 Darmstadt, Germany}

\email{\authormark{*}cristofero.oglialoro@gmx.de}

\begin{abstract*}
Developing advanced technologies for sensing and imaging biological samples is crucial for medical applications, making quantum-enhanced methods particularly valuable, as they promise significant benefits over classical techniques. 
An important aspect of biological imaging is the characterization of tissue, which often involves resolving complex structural information such as birefringence and diattenuation. 
These measures require polarization-sensitive sensing which remains largely unaddressed in quantum-imaging techniques with undetected light. 
However, the bicolor nature and supreme phase sensitivity of nonlinear interferometers make them particularly advantageous for biological sensing.
Hence, we theoretically introduce controllable polarizations of the interrogating light in a quantum-imaging setup and show the potential of nonlinear interferometers to simultaneously sense birefringence and diattenuation with undetected light while discussing both the low- and high-gain regime.
\vspace{.5em}
\newline
This article has been published in \href{https://doi.org/10.1364/OE.560824}{Optics Express \textbf{33}, 17 (2025)} under the terms of the \href{https://creativecommons.org/licenses/by/4.0/}{Creative Commons Attribution 4.0 License [CC BY]}.
\end{abstract*}

\section{Introduction}
Quantum imaging with undetected photons~\cite{barretolemos2014, lahiri2015} of biological specimens represents a particularly attractive technology, since it utilizes its bicolor property~\cite{kutas2020a, kutas2025} for chemically selective infrared imaging~\cite{diem2013, wrobel2018} at low intensities to overcome issues of phototoxicity~\cite{kasprzycka2024}.
However, such biological samples often exhibit birefringence and diattenuation~\cite{mehta2013a, mann2024}, properties which characterize a specimen but are so far not accounted for in most quantum-imaging schemes.
In this article, we theoretically demonstrate that nonlinear interferometers (NLIs)~\cite{chekhova2016} are capable of simultaneously sensing both properties with undetected light and develop schemes to extract this information.

NLIs perform squeezing-assisted quantum imaging where the interrogating photon is never detected~\cite{barretolemos2014, chekhova2016, fuenzalida2024a}.
So far, they have been employed in several applications, including quantum holography~\cite{topfer2022, topfer2025,leon-torres2024, pearce2024}, spectroscopy~\cite{kalashnikov2016, lindner2020, hashimoto2024}, and optical coherence tomography~\cite{valles2018, paterova2018}.
The technology is based on parametric down-conversion (PDC), a process which nonlinearly couples three different modes (pump, signal, and idler) of possibly different polarization, while being inherently sensitive to the polarization of the inducing or seeding radiation~\cite{kwiat1995}. 
Therefore, NLIs are in principle susceptible to different light polarizations, a property which is, however, rarely utilized in the context of sensing with undetected light~\cite{fuenzalida2024a, arya2025}.

Since the interferometers are based on the interference of photons of two PDC processes, modulation of the light polarization between the first and second nonlinearity can introduce distinguishability, and thus diminish the interference. 
This may be countered by employing quantum erasure~\cite{herzog1995, gemmell2024}, but also may be used to nondestructively determine a quantum state~\cite{fuenzalida2024b} or to infer the polarization rotation induced by an optical element, such as the Faraday rotation imprinted by a crystal subject to a magnetic field in order to deduce the Verdet constant~\cite{chakraborty2025}.

Although polarization effects have been included in NLIs to sense the retardation of a birefringent sample~\cite{paterova2019},  they have primarily been applied to objects with polarization-independent light attenuation, e.\,g., absorption.
Real-life samples may however feature diattenuation along with birefringence~\cite{mehta2013a, mann2024}, which cause polarization-dependent light attenuation and a polarization-dependent phase retardation, respectively.
Such features reveal structural details of biological samples, where each property may emphasize distinct tissue regions~\cite{menzel2017}.
For this reason, diattenuation and birefringence are employed in the polarimetric characterization of biological samples, e.\,g., the tracking of cancer cell deaths over time after being subjected to chemotherapy medication~\cite{fernandez-perez2019} or the distinction between cancerous and healthy tissue~\cite{du2014}.
Conventional techniques imaging birefringent and diattenuating samples are polarimetry with laser light~\cite{he2021, louie2022} or X-ray radiation~\cite{palmer2014, grabiger2020}, polarimetric ghost imaging~\cite{chirkin2018, hannonen2020, vega2021}, and (quantum) optical coherence tomography~\cite{deboer1997, booth2004, sukharenko2024}.
These all require the detection of photons interacting with the sample and generally lack the feature of bicolor sensing.

While many biological specimens typically exhibit birefringence over a broad spectral range, including the visible~\cite{mehta2013a, mann2024, fernandez-perez2019}, we propose a versatile setup which supports bicolor imaging, in which the sample is illuminated at a wavelength different from that of detection, which is performed with more efficient detectors such as silicon-based devices.
This capability is particularly beneficial for label-free, chemically selective, polarization-sensitive absorbance detection of samples with these characteristics at mid-infrared frequencies, enabling, for example, the analysis of structural and orientational features in samples such as collagen fibers in human bone marrow tissue~\cite{mankar2022}.
Some samples, such as the corpus callosum connecting the left and right cerebral spheres in placental mammal brains, even display strong birefringence in the low terahertz range~\cite{chernomyrdin2023}.
Since birefringence is in part caused by scattering effects, different spectral ranges probe distinct length scales within the specimen’s structure, enabling access to complementary structural information.
Similarly, additional optical resonances, such as vibrational resonances at specific wavelengths, can be used to obtain structural information about the specimen.
For inorganic or synthetic materials, the deep penetration depth and non-ionizing nature of terahertz radiation make bicolor imaging an attractive alternative for birefringence characterization across a range of crystals and metamaterials~\cite{wiesauer2013}.
Evidently, in addition to its applicability for sensing biological samples, the presented setup is also well-suited for characterizing optical elements such as waveplates and birefringent crystals.

For classical schemes, the detection of probing light at mid-infrared and terahertz frequencies remains challenging.
Recent advancements in nonlinear interferometry have demonstrated a probing of samples at mid-infrared wavelengths while detecting signals at near-infrared frequencies~\cite{kviatkovsky2020}, or to probe at terahertz  frequencies while detecting visible light~\cite{kutas2020a, kutas2025}.
Hence, such an approach circumvents the need to detect mid-infrared or terahertz frequencies directly, while retaining their sensing properties.
Accordingly, we propose a scheme capable of sensing over a broad frequency spectrum using undetected light, thereby enhancing its applicability to a wide variety of biological and inorganic specimens.

In this setup, the diattenuating and birefringent sample is probed by idler light generated in a pumped nonlinear crystal via PDC.
This idler, after interrogating a sample, along with a reference signal photon of possibly different frequency, seeds a second PDC process which again may produce or annihilate signal and idler photon pairs.
The idler seed induces coherence and leads to first-order interference in the signal mode~\cite{yurke1986, wang1991, wiseman2000}.
Due to the nonlinear mixing, the signal light carries all sample information without ever interacting with it while utilizing bicolor sensing.
The proposed setup constitutes a Mach-Zehnder-type nonlinear interferometer.
Compared to the widely employed Michelson-type interferometer~\cite{chekhova2016, paterova2019}, this configuration introduces additional experimental challenges, primarily due to the need for precise alignment of the two crystal outputs and the associated stability.
However, the Mach-Zehnder configuration provides greater flexibility for polarization control of the probing light, since it is not constrained by the symmetry requirements.
Previous work~\cite{paterova2019} used polarization effects in a Michelson-type nonlinear interferometer to determine the retardance of a waveplate with an accuracy of up to $\pm 2 \times 10^{-3} \pi$.
Even when using the more stable configuration, the measurement uncertainty  was dominated by phase drifts within the interferometer and fluctuations of the photocount rate of the detector.
Our scheme requires phase scans between the two polarization components in the idler arm, necessitating additional spatial separation and the use of piezo-translators or spatial light modulators, which introduces added experimental complexity.
When applied to sensing biological tissue, we anticipate comparable or larger uncertainties than the characterization of the phase retarder~\cite{paterova2019}, as such specimens can introduce further technical noise due to scattering and inhomogeneities.
Hence, these limitations can be considered a lower bound for experiments performed with our proposed Mach-Zehnder-type setup.
But even in the absence of technical noise, measurements remain inherently shot-noise limited at low pump intensities.

Even though biological samples are generally phototoxic, increasing the gain beyond the spontaneous regime may lead in SU(1,1) interferometer-type configurations~\cite{yurke1986} to an enhanced sensitivity below the classical shot-noise limit and can even approach the ultimate Heisenberg limit~\cite{giese2017, schaffrath2024a}.
Although sub-shot-noise sensing of inhomogeneous biological samples is not expected in this work due to considerable loss channels induced by the optical elements and polarization-dependent sample characteristics, large squeezing may still offer an increased sensitivity compared to conventional birefringence and diattenuation sensing methods.

To infer the complex sample properties of interest with an SU(1,1) interferometer, theoretically introduced in Sec.~\ref{ch:setup}, we propose to include additional waveplates to probe both phase-retarding and diattenuating axes of an oriented sample in Sec.~\ref{ch:output}, which results naturally in three-mode interference.
We present different strategies to deduce both diattenuation and birefringence properties of a sample from the features of this interference and transfer the results to the high-gain regime. 
Finally, we generalize our results to the sensing of samples of unknown orientation in Sec.~\ref{ch:rotation}.
In Appendices~\ref{ch:transformations} and~\ref{ch:WP} we present the transformations modeling the interferometer and include in Appendix~\ref{ch:sample rotation} the rotation of a sample as an additional parameter.

\section{Nonlinear interferometer with birefringent, diattenuating sample \label{ch:setup}}

\begin{figure}
    \centering   \includegraphics[width=0.7\linewidth]{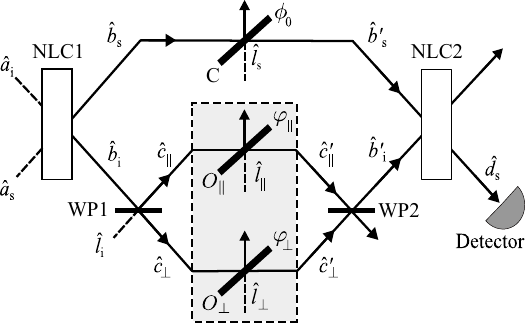}
    \caption{Nonlinear interferometer consisting of two nonlinear crystals NLC1 and NLC2 (that generate nondegenerate signal and idler photons), waveplates WP1 and WP2 (that are used to tune the polarization of the idler arm), and a beam splitter C (which controls the signal arm).
    The diattenuating and birefringent sample (described by the action of two transformations $O_\pa$ and $O_\pp$ for each polarization component) is sensed by the idler photons, while the signal photons are measured by a detector after the second crystal. All transformations describing the interferometer are stated in Appendix~\ref{ch:transformations}.}
    \label{fig:setup}
\end{figure}

The setup drawn in Fig.~\ref{fig:setup} depicts a nonlinear SU(1,1) interferometer~\cite{yurke1986}, which consists of two nonlinear crystals (NLC1 and NLC2) pumped by coherent laser light (not shown). 
The unseeded NLC1 has only vacuum input modes, described by bosonic annihilation operators $\hat{a}_\s$ and $\hat{a}_\i$, and generates nondegenerate signal and idler photons ($\hat{b}_\s$ and $\hat{b}_\i$), whose different wavelengths can be used for bicolor sensing~\cite{kutas2020a, kutas2025}. 
The input and output modes are linked through a Bogoliubov transformation, as detailed in Appendix~\ref{ch:transformations}.
After the generation of these two beams, only the idler photons impinge on a birefringent and diattenuating sample (represented by a gray box). 
A waveplate (WP1) prior to the object ensures an elliptical polarization of the idler so that both polarizations of the object may be probed, which is schematically shown as a splitting of the idler path into a perpendicular ($\hat{c}_\perp$) and a parallel ($\hat{c}_\parallel$) component.
In reality, both polarizations populate the same spatial mode.
Vacuum noise of the initially unpopulated polarization component described by $\hat{l}_\i$ couples into the idler at WP1.
The effect of the sample on each polarization component is described by a beam splitter transformation $O_\perp$ and $O_\parallel$, which imprint phases $\varphi_\pp$ and $\varphi_\pa$ and induce loss associated with the vacuum input modes $\hat{l}_\pp$ and $\hat{l}_\pa$.
Hence, these two transformations include both diattenuation and birefringence.
A second WP2 then maps the phase and intensity information onto the original idler polarization inducing coherence ($\hat{b}'_\i$).
Since the orthogonal polarization component is distinguishable from the initially created photons, it causes no interference effects and is subsequently traced out.
We are primarily focused on the effects of diattenuation and birefringence, therefore we assume in Fig.~\ref{fig:setup} that the sample is oriented such that the initial idler polarization matches one axis of the sample. We discuss the scenario of a rotated sample in Sec. \ref{ch:rotation}.

\begin{figure}
    \centering   \includegraphics[width=0.7\linewidth]{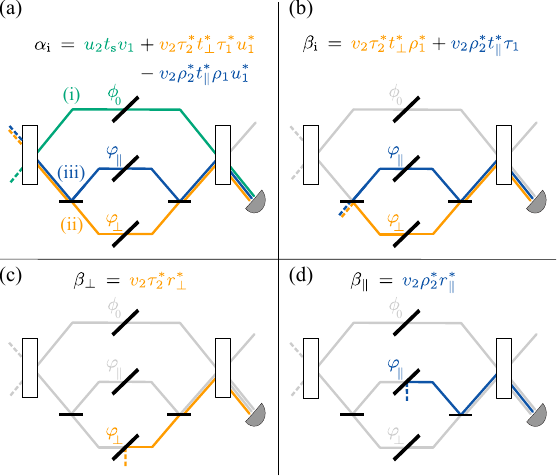}
     \caption{Visualization of different paths through the interferometer and the corresponding coefficients that contribute to the output signal mode $\hat{d}_\s$.
     The coefficient $\alpha_\i$ is a three-path superposition of the signal (green) and idler (blue, orange) photons (a), while $\beta_\i$ describes a superposition of two noise contributions induced by WP1 (b).
     The coefficients $\beta_{\pp,\pa}$ are associated with the paths of the noise induced for each polarization of the sample (c-d).
     The signal coefficients $\alpha_\s=u_2 t_\s u_1 +v_2\tau^*_2t^*_\pp \tau^*_1v^*_1-v_2\rho^*_2 t^*_\pa\rho_1v^*_1$ and $\beta_\s=u_2 r_\s$ do not contribute to the interference signal Eq.~\eqref{eq:N} since they are associated with the annihilation operators $\hat{a}_\s$ and $\hat{l}_\s$ in the output mode $\hat{d}_\s$ and are therefore not displayed.}
    \label{fig:3paths}
\end{figure}

On the other arm of the interferometer, the signal beam can be controlled by a beam splitter (C) that shifts the phase $\phi_0$ of the signal mode ($\hat{b}'_\s$)  or can be used as a block.
The modified signal ($\hat{b}'_\s$) and idler ($\hat{b}'_\i$) modes seed NLC2, which induces another nonlinear process and mixes them.
This ensures that the phase information is encoded in the joint quantum state of both frequency components, even though the signal photon never interacted with the sample.
This way, coherence induced by the idler photons can be observed in the signal mode.
Finally, only the photon number in the signal output mode $\hat{d}_\s$ of NLC2 is detected, which highlights the main goal of bicolor sensing.
It is connected to the input operators that fulfill the usual bosonic commutation relations 
$[\hat a_j,\hat a_k^\dagger]=\delta_{jk}=[\hat l_j,\hat l_k^\dagger]$ and 
$[\hat d_\s,\hat d_\s^\dagger]=1$, while all other combinations vanish, and takes the form
\begin{align}
\begin{split}
\label{eq:d_s}
 \hat{d}_\s=\alpha_\s \hat{a}_\s+\alpha_\i \hat{a}^\dagger_\i+\beta_\s \hat{l}_\s+\beta_\i \hat{l}^\dagger_\i+\beta_\pp \hat{l}^\dagger_\pp+\beta_\pa \hat{l}^\dagger_\pa.
\end{split}
\end{align}
The observed pattern is caused by the interference of three possible paths through the interferometer, specifically the signal path and both polarization components of the idler path.
This path information is included in the complex coefficients $\alpha_{\s,\i}$, $\beta_{\s,\i}$ and $\beta_{\pp,\pa}$ from Eq.~\eqref{eq:d_s} and is visualized in Fig.~\ref{fig:3paths}, where the explicit form of the coefficients is given.
They are obtained from the transformations listed in Appendix~\ref{ch:transformations} and include the complex coefficients $u_{j}$ and $v_{j}$ associated with the Bogoliubov transformation induced by NLC$j$, which are connected by $|u_{j}|^2-|v_{j}|^2=1$.
The phase of the pump at the respective crystal is included in the phase of $v_j$ while $|v_{j}|^2$ corresponds to the number of photons generated in each mode.
The complex transmission and reflection coefficients $\tau_{j}$ and $\rho_{j}$ describing the action of WP$j$ satisfy the conditions of a SU(2) beam splitter-type transformation.
They can be adjusted for either a half-waveplate (HWP) or quarter-waveplate (QWP) oriented at any given angle.
Further details on these coefficients are provided in Appendix~\ref{ch:WP}.
For the diattenuating and birefringent sample, the transmission (reflection) coefficients  $t_p$ ($r_p$) associated with each polarization component $p=\pp,\pa$ account for loss and impart a phase $\varphi_p$ on the respective polarization component. 
Similarly, the transmission coefficient $t_\s$ of the beam splitter C encodes a phase shift $\phi_0$ on the signal arm.
We associate these coefficients with the probability for propagating along a particular path through the interferometer and explicitly discuss $\alpha_\i$ from Fig.~\ref{fig:3paths}(a), as it accounts for the main contribution:
In path (i, green), signal photons generated in NLC1 are controlled by C and seed NLC2.
In path (ii, orange) a fraction of idler photons generated in NLC1  senses the $\pp$ polarization of the sample, is mixed with the $\pa$ polarization mode, and then seeds the signal output of NLC2.
In path (iii, blue) the complementary fraction of idler photons transformed by WP1 probes the $\pa$ polarization, is transformed back to the initial polarization by WP2 and then seed the PDC process of NLC2.
The remaining coefficients have their origin in loss modes of the respective optical elements and are seeded with vacuum.
This three-path feature may be used to distinguish the contributions of both polarizations in the interferometer output.

The mode operator $\hat{d}_\s$ covering these traits is applied to the vacuum state $\ket{0}$, which results in
\begin{align}
\hat{d}_\s\ket{0}=\alpha_\i\ket{1_\i}+\beta_\i \ket{1_{l\i}}+\beta_\pp\ket{1_{l\pp}}+\beta_\pa\ket{1_{l\pa}}.
\end{align}
Each term corresponds to the creation of a photon in the respective mode indicated by $\ket{1_k}$, while the rest remain in vacuum.
The photon number $N$ detected in the signal path is therefore
\begin{align}
N=\bra{0}\hat{d}^\dagger_\s\hat{d}_\s\ket{0}
= |\alpha_\i|^2 + |\beta_\i|^2 + |\beta_\pp|^2 +|\beta_\pa|^2.\label{eq:N}
\end{align}
We interpret the term $|\alpha_\i|^2$ as a three-path interference of the green, blue and orange paths in Fig.~\ref{fig:3paths}(a).
The contribution $|\beta_\i|^2$ describes the interference of the two noise channels induced by WP1.
The coefficients $|\beta_\pp|^2$ and $|\beta_\pa|^2$ include the amplitudes of the noise contributions caused by each polarization of the sample.

\section{Interferometer output \label{ch:output}}
In this section, we evaluate the observed signal from Eq.~\eqref{eq:N} in the low-gain regime and provide a measurement strategy to extract the phase and transmission information of a diattenuating and birefringent sample inserted in the idler arm.
Furthermore, we transfer our results to the high-gain regime and discuss deleterious effects caused by idler interferences.

\subsection{Low-gain interference \label{ch:lg}}
So far most quantum imaging experiments were performed for low pumping intensities where the probed objects are illuminated by a weakly squeezed vacuum state, \text{i.\,e.}, single photon pairs  \cite{lahiri2019, fuenzalida2024a, kopf2025}. 
Since the generated photon number $|v_{1}|^2$ increases exponentially with the pump intensity, our results provide this scenario when all terms quadratic in the generated photon number are neglected. 
We assume equal pumping of both crystals \cite{manceau2017, giese2017, manceau2017a} $|v_{1}|^2=|v_{2}|^2=V$ and evaluate the modulus squared of the coefficients from Eq.~\eqref{eq:N} to find the low-gain result
\begin{align}
\begin{split}
    N_\text{lg}=V\big[ |t_\s|^2+ 1+2|t_\s\tau_2 t_\pp\tau_1|\cos\left(\phi_\tau+\phi_0+\varphi_\pp\right)
    -2|t_\s\rho_2 t_\pa \rho_1|\cos\left(\phi_\rho+\phi_0+\varphi_\pa\right)\big],\label{eq:Nlg}
\end{split}
\end{align}
where $\phi_\tau$ and $\phi_\rho$ are phases that are adjustable by rotating the WPs (see Appendix \ref{ch:WP}).
The phase $\phi_0$ includes the phase shift of the signal by C and, effectively, the phase difference of the pump laser between the crystals.
The three-path interference amounts to a beating of two competing interference patterns which oscillate with the phases associated with each polarization $\varphi_\pp$ and $\varphi_\pa$.
Next, we discuss the diattenuating and birefringent sample in terms of its differential and mean optical characteristics. 
For that, we define the mean phase $\bar{\varphi} = (\varphi_\pp + \varphi_\pa)/2$ and mean transmission coefficient  $\bar{t}=|\tau_2 t_\pp \tau_1|+|\rho_2 t_\pa \rho_1|$ as well as the retardance (associated with birefringence) $\delta \varphi = \varphi_\pp - \varphi_\pa$
and diattenuation $\delta t=2\left(|\tau_2 t_\pp \tau_1|-|\rho_2 t_\pa \rho_1|\right)$.
With these definitions, the photon number is given by
\begin{align}
\begin{split}
    N_\text{lg}=\frac{A}{2}\bigg[ 1 + \mathcal{V}_\delta \cos\left(\frac{\delta \varphi + \delta \Phi}{2}\right)\cos\left(\bar{\varphi}+\bar \Phi\right)- \bar{\mathcal{V}} \sin\left(\frac{\delta \varphi + \delta \Phi}{2}\right)\sin\left(\bar{\varphi}+\bar \Phi\right)\bigg] \label{eq:beating}
\end{split}
\end{align}
with the mean and differential phases of the interferometer $\bar{\Phi} = \phi_0 + (\phi_\tau + \phi_\rho)/2$ and $\delta \Phi = \phi_\tau - \phi_\rho$.
We identify the amplitude of the beating pattern 
\begin{subequations}
\begin{equation}
    A=2V\left(|t_\s|^2+1\right)
\end{equation}
and find two independent visibilities
\begin{equation}
    \mathcal{V}_\delta=\frac{|t_\s|\,\delta t}{|t_\s|^2+1}
    ~\text{ and }~
    \bar{\mathcal{V}}=\frac{2|t_\s|\,\bar{t}}{|t_\s|^2+1}.
\end{equation}
\end{subequations}
These convey the maximum visibilities achievable for a measurement of the differential and mean transmittance, respectively.

\begin{figure}
    \centering
    \includegraphics[width=0.7\linewidth]{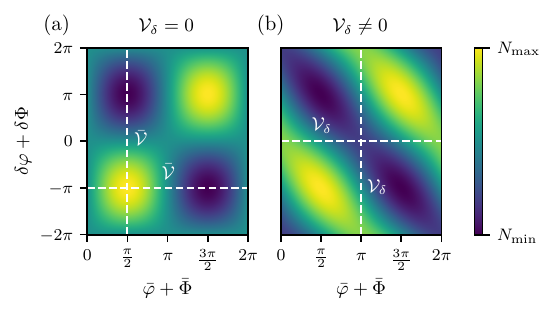}
        \caption{Low-gain interference signal from Eq.~\eqref{eq:beating} depicted for (a) polarization-independent attenuation ($|t_\pp| = |t_\pa| = 0.9$) and (b) diattenuation ($|t_\pp| = 0.9,\ |t_\pa| = 0.2$). The white dashed lines indicate fixed phase settings for which the visibilities $\bar{\mathcal{V}}$ and $\mathcal{V}_\delta$ can be inferred from a scan of the other phase. Here, we assume no photon loss in the signal arm, i.\,e., $|t_\s|=1$}.
    \label{fig:beating}
\end{figure}

The photon number from Eq.~\eqref{eq:beating} exhibits a beating between contributions from the differential and mean transmission coefficients.
To our knowledge, only equal transmittance for both polarization components has been studied in the context of quantum imaging with undetected photons, such as the signature induced by Faraday rotation~\cite{chakraborty2025} onto the observed visibility. 
Fig.~\ref{fig:beating}(a) illustrates the simplified case without diattenuation where $\mathcal{V}_\delta = 0$, \text{i.\,e.}, $\delta t = 0$.
In this case, the interference pattern can be described in terms of an effective visibility $\bar{ \mathcal{V}} \sin [(\delta \varphi + \delta \Phi)/2]$ which includes differential phase information between both polarizations.
These oscillations exhibit the visibility $\bar{\mathcal{V}}$ for $\delta \varphi+\delta \Phi=\pm\pi$ and $\bar {\varphi}+\bar{\Phi}=\pi/2,\, 3\pi/2$, as indicated by the white dashed lines in Fig.~\ref{fig:beating}.
This visibility can be retrieved by alternatively adjusting either of the two phase settings and scanning the other phase over one (half) period.
At $\delta \varphi+\delta \Phi=0$ we observe no interference.
This phase setting can be achieved by removing the sample and applying two QWPs ($\theta=\pi/2$) at orientations of $\gamma=\pi/4$, which polarizes the idler perpendicularly to its original orientation.
This effect causes distinguishability and prevents the idler from inducing coherence in the signal mode.
Different WP settings and their effects on the observed visibility have been used, for example, to determine the retardance of a WP at different frequencies~\cite{paterova2019}, to employ quantum erasure~\cite{herzog1995, gemmell2024}, or for quantum state tomography \cite{fuenzalida2024b}, where the polarization state of a photon can be inferred without directly measuring it.

Due to the relevance of diattenuation for biological samples~\cite{du2014, menzel2017, fernandez-perez2019, dubreuil2020}, we extend this discussion to the anisotropic attenuation of the polarizations ($\mathcal{V}_\delta\neq0$).
We depict an exemplary two-dimensional interference pattern in Fig.~\ref{fig:beating}(b).
It shows a tilting and smearing of the interference pattern and isolates the visibility $\mathcal{V}_\delta$ at $\delta \varphi+\delta \Phi=0$ and $\bar {\varphi}+\bar{\Phi}=\pi$.
The phase settings along which $\bar{\mathcal{V}}$ may be obtained remain unchanged.  
In the case of diattenuation, both visibilities can be obtained by phase scans along the corresponding dashed lines.
However, such scans require precise knowledge of the mean or differential interferometer and object phases. 
We therefore employ a Fourier analysis of the beating pattern as described in Sec.~\ref{ch:Fourier}.

\subsection{Fourier analysis of beating  \label{ch:Fourier}}
To perform a Fourier analysis of the beating pattern from Eq.~\eqref{eq:beating}, we change the phase settings of subsequent measurements over time with rates $\dot{\phi}_0$ and $\delta\dot{\phi}$.
The signal arm is  altered by scanning the beam splitter C as \red{$\phi_0 \rightarrow \bar{\xi} +\dot{\phi}_0 t$}.
Simultaneously, a phase scan between both polarizations of the idler arm is performed by adding the phase $\delta \xi+ \delta\dot{\phi} t$.
One option is to include polarizing beam splitters after the object to spatially separate the polarization components, which can then be phase shifted individually before being spatially recombined.
Alternatively, additional retarders can be inserted to introduce polarization dependent phases.  
However, a phase scan of the retardance $\delta \varphi$ requires knowledge of the sample orientation.
We therefore discuss in Sec.~\ref{ch:rotation} a sample of unknown orientation and present a scheme to extract this information. 

In order to separately scan the retardance $\delta \varphi$ and mean phase $\bar{\varphi}$ the average phase shift of the idler arm must vanish.
Since the precise phase at $t=0$ is generally unknown, we include offsets $\bar{\xi}$ and $\delta \xi$ in our description.
As a result, the interferometer needs to be calibrated first.
To that end, the object is removed and two QWPs ($\theta=\pi/2$) are applied at orientations of $\gamma_1=\pi/4$ and $\gamma_2=3\pi/4$, respectively.
Moreover, we assume $|t_\s| =1$ so that no photons are lost on the signal arm.
Without sample, the signal therefore reduces to
\begin{align}
    N_0(t) = 2V\left[1+\cos\left(\frac{\delta \xi + \delta\dot{\phi} t}{2}\right)\cos(\bar{\xi} +\dot{\phi}_0 t)\right].
\end{align}
By scanning the signal and idler arms separately, the phase offsets $\bar{\xi}$ and $\delta \xi$ can be inferred.
Next, the sample may be inserted for the full time dependent interference signal
\begin{align}
\begin{split}
    N_\text{lg}(t)=\frac{A}{2}\bigg[& 1 - \mathcal{V}_\delta \sin\left(\frac{\delta \varphi+\delta \xi + \delta\dot{\phi} t}{2}\right)\sin\left(\bar{\varphi}+\bar{\xi}+ \dot{\phi}_0 t\right)\\
    &+ \bar{\mathcal{V}} \cos\left(\frac{\delta \varphi+\delta \xi + \delta\dot{\phi} t}{2}\right)\cos\left(\bar{\varphi}+\bar{\xi}+ \dot{\phi}_0 t\right)\bigg]. \label{eq:beating2}
\end{split}
\end{align}
Transforming this signal to Fourier space $\Omega$ for equal scanning rates of signal and idler arm ($\dot{\phi}_0=\delta\dot{\phi}=\omega_\text{scan}$) and identifying $\bar{\mathcal{V}} - \mathcal{V}_\delta =|t_\pa|$ and $\bar{\mathcal{V}} + \mathcal{V}_\delta=|t_\pp|$, we find
\begin{align}
\begin{split}
\frac{ 4 \tilde{N}_\text{lg}(\Omega)}{\sqrt{2\pi}A}=2\delta(\Omega)+&|t_\pp|\left[\kappa_+\, \delta(3 \omega_\text{scan} + 2 \Omega)+ \kappa_-
 \, \delta(3 \omega_\text{scan} - 2 \Omega)\right]\\
 +& |t_\pa|\left[\ve_+\, \delta(\omega_\text{scan} + 2 \Omega)+\ve_-
 \, \delta(\omega_\text{scan} - 2 \Omega)\right]\label{eq:fourier}
\end{split}
\end{align}
with the phase factors
\begin{align}
    \ve_{\pm} = \text{exp}\left\{ \pm\i \left(\bar\varphi+\bar\xi-\frac{\delta\varphi+\delta\xi}{2} \right)\right\}
    ~\text{ and }~
    \kappa_{\pm} =  \text{exp}\left\{ \pm\i \left(\bar\varphi+\bar\xi+\frac{\delta\varphi+\delta\xi}{2} \right)\right\}.
\end{align}
We identify five frequency components $\Omega = 0$, $\Omega = \pm\omega_\text{scan}/2$ and $\Omega = \pm3\omega_\text{scan}/2$ of the beating pattern.
The height of the peak at $\Omega = 0$ retrieves the amplitude $A$, while the displaced peaks can be directly used to read-off the transmission coefficients $|t_\pa|$ at $\Omega = \omega_\text{scan}/2$ and $|t_\pp|$ at $\Omega = 3\omega_\text{scan}/2$.

The phases of the peaks in the Fourier transform Eq.~\eqref{eq:fourier} can be used to infer the mean phase
\begin{subequations}
\begin{align}
  \bar{\varphi}+\bar{\xi} =\frac{1}{2}\left(\arctan\left[\frac{\Im\, \kappa_+}{\Re\, \kappa_+}\right]+ \arctan\left[\frac{\Im\, \ve_+}{\Re\, \ve_+}\right] \right)
\end{align}
and the retardance
\begin{align}
   \delta\varphi+\delta\xi = \arctan\left[\frac{\Im\, \kappa_+}{\Re\, \kappa_+}\right]- \arctan\left[\frac{\Im\, \ve_+}{\Re\, \ve_+}\right] 
\end{align}
\end{subequations}
with known offsets $\delta\xi$ and $\bar{\xi}$. 
Therefore, a Fourier transform of the time dependent 
\cite{footnote1}
complex beating pattern from Eq.~\eqref{eq:beating2} facilitates a straightforward separation of the contributions from the sample's axes. 
Due to the factor of 1/2 for the differential phase scan, it is preferable to scan both interferometer arms with the same rate $\dot{\phi}_0=\delta\dot{\phi}=\omega_\text{scan}$  for an optimized separation of the peaks in Fourier space.

Thus, the presented method enables the detection of the diattenuation and phase information of an oriented birefringent and diattenuating object.
\begin{figure}
    \centering
    \includegraphics[width=0.7\linewidth]{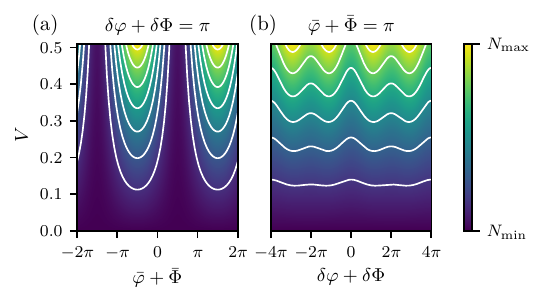}
    \caption{Observed interference signal in the high-gain regime obtained from Eq.~\eqref{eq:Nhg} for increasing gain and a birefringent, diattenuating sample.
    The transmission coefficients for both polarizations are $|t_\pp| = 0.9,\ |t_\pa| = 0.8$.
    We observe different oscillatory behavior proportional to the mean phase $\bar{\varphi}$ for $\delta \varphi+\delta\Phi=\pi$ (a) and proportional to the retardance $\delta\varphi$ for $\bar{\varphi}+\bar{\Phi}=\pi$ (b). The latter exhibits a complex pattern which gets more pronounced for increasing gain. Here, we assume no photon loss in the signal arm, i.\,e., $|t_\s|=1$.}
    \label{fig:hg}
\end{figure}

\subsection{Operation in the high-gain regime\label{ch:hg}}
Our description of the setup also includes the operation in the high-gain regime. 
The physical consequence of this regime is an increased squeezing of the generated light fields and therefore an increase of the photon number of the participating signal and idler modes.
For that, we also include contributions that scale with $V^2$ to find the photon number in the high-gain regime
\begin{align}
\begin{split}
    N_\text{hg}=N_\text{lg} (1+V) - V^2 +\frac{V^2\delta t^2}{4}\cos^2\left(\frac{\delta \varphi+ \delta \Phi}{2}\right)+V^2\bar{t}^2\sin^2\left(\frac{\delta \varphi+ \delta \Phi}{2}\right)\label{eq:Nhg}\,
\end{split}
\end{align}
where we included the low-gain photon number $N_\text{lg}$  from Eq.~\eqref{eq:beating}.
To enable a quantitative comparison of the low- and high-gain results, we consider the evolution of the interference pattern as a function of $\bar{\varphi}$, while keeping $\delta \varphi+ \delta \Phi=\pi$ fixed and observe
\begin{align}
\begin{split}
\left.N_\text{hg}\right|_\pi=V\left(|t_\s|^2+1\right)+V^2\left(|t_\s|^2+\bar{t}^2\right)-2|t_\s|\bar{t}\,V\left(1+V\right)\sin\left(\bar{\varphi}+\bar \Phi\right).
\end{split}
\end{align}

This interference pattern is depicted in Fig.~\ref{fig:hg}(a) for increasing gain.
In this case the effective high-gain visibility for a $\bar{t}$ measurement is hence always larger or equal than the corresponding low-gain visibility
\begin{align}\left.\mathcal{V}_\text{hg}\right|_\pi = \frac{2|t_\s|(1+V)\bar{t}}{1+|t_\s|^2(1+V)+\bar{t}^2V} \geq \bar{\mathcal{V}},
\end{align}
which has been already observed theoretically and experimentally for an isotropic phase object ~\cite{michael2021}. 
The evolution of the high-gain interference pattern from Eq.~\eqref{eq:Nhg} with the retardance $\delta \varphi$ is more complex, since it additionally depends on higher harmonic contributions (last two terms). 
Thus, the visibility is significantly modified for an estimation of $\delta \varphi$ when  compared to the low-gain.
This effect is shown in Fig.~\ref{fig:hg}(b). 
Compared to the mean phase $\bar{\varphi}$, the photon number exhibits a complex beating behavior as a function of the retardance $\delta \varphi$.

\begin{figure}
        \centering
        \includegraphics[width=0.7\linewidth]{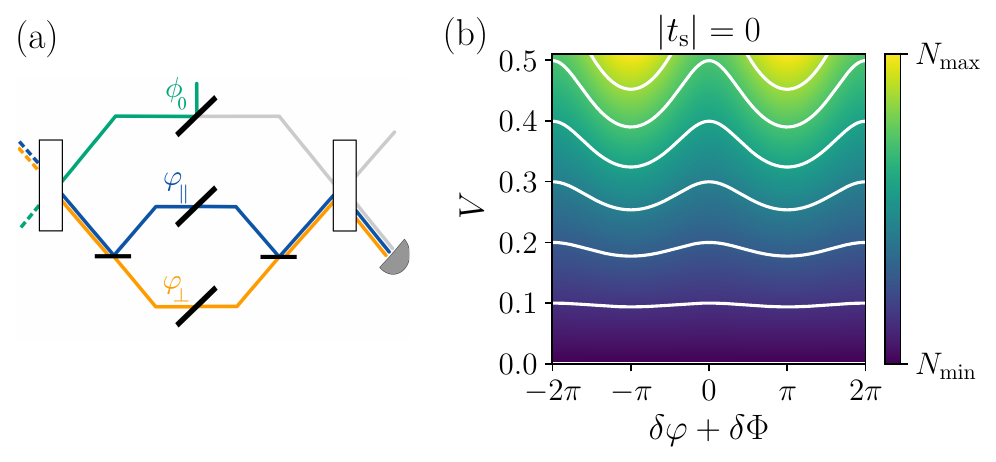}
        \caption{High-gain interferometer (a) where the signal beam between both crystals is blocked.
        Nevertheless, interference between two idler paths can be observed by varying $\delta \varphi$.
        While vanishing in the low-gain regime ($V\ll1$), the amplitude of these oscillations increases with gain (b).
        The combination of WP2 and NLC2 can be interpreted as a nonlinear bicolor analyzer and as an indicator for the high-gain regime.
        For this plot, we chose the transmission coefficients $|t_\pp| = 0.9,\ |t_\pa| = 0.8$.}
        \label{fig:hg_blocked}
\end{figure}

Interestingly, the higher-harmonic contributions are even present if we block the signal beam between both crystals.
In fact, for $|t_\s| = 0$ without any assumptions on the phases, we observe the blocked interference pattern
\begin{align}
    N_\text{b}=V +  V^2\bigg[\frac{\delta t^2}{4}\cos^2\left(\frac{\delta \varphi+ \delta \Phi}{2}\right)+\bar{t}^2\sin^2\left(\frac{\delta \varphi+ \delta \Phi}{2}\right)\bigg].\label{eq:blocked signal3}
\end{align}
Since these oscillations scale quadratically with $V$ and by that only appear in the high-gain regime, they can be attributed to polarization-dependent interference of the idler photons on WP2, which in turn modulates the seed of NLC2.
A similar attenuation of the output intensity by the transmittance for both polarizations has been observed in classical polarized light microscopy~\cite{mehta2013a}.
Similar to our setup, the attenuation of the probing light intensity  was also governed by higher harmonics.
However, our results differ in that the higher-order attenuation occurs at frequencies that are never detected, so that the combination of WP2 and NLC2 can be interpreted as a nonlinear, bicolor analyzer. The detected photons are generated by pumping the second crystal and parametrically converting the interrogating idler photons to this frequency, potentially allowing the setup to function as a quantum frequency transducer~\cite{wang2022, han2023} or an up-conversion imaging system~\cite{barh2019, huang2022} used in biological sensing~\cite{junaid2019}. 
This insight highlights the versatility of the proposed setup, as it can be readily transformed from a nonlinear quantum imaging setup with undetected light in the high-gain regime to an up-conversion imaging system seeded by one mode of a two-mode squeezed vacuum state, modulated by the object.
However, the performance of this up-conversion scheme is limited by the conversion rate of the first crystal and must be compared to that achieved with laser light at the idler frequency used as a seed of NLC2.
Since our measurement is based on intensity detection, we expect noise arising from the photon statistics of the seed, which is thermal.
Because one mode is discarded, we therefore anticipate no advantage over a direct up-conversion scheme using laser light.

In Fig.~\ref{fig:hg_blocked}(a) we show the setup leading to the blocked interference signal Eq.~\eqref{eq:blocked signal3}.
In analogy to the discussions in Sec.~\ref{ch:setup}, we perceive this as two-path interference of the orange and blue paths.
In Fig.~\ref{fig:hg_blocked}(b) we show Eq.~\eqref{eq:blocked signal3} depending on the gain.
For increasing pump intensity, the high-gain oscillations start to dominate over the low-gain baseline, which is phase independent due to the block.
Hence, the observation of oscillations can serve as an indicator for the high-gain regime.

Naturally, even when $t_s \neq 0$, this interference effect plays a crucial role in the observed interference pattern, see Fig.~\ref{fig:hg}.
In this case, the signal mode in NLC2 is seeded, while the idler-mode seed experiences strong modulation due to the interference of both polarization components.
In fact, our results in the high-gain regime show that the low-gain result is not merely amplified by a factor of $(1+V)$; instead, the up-conversion signal $N_\text{b}$ is also added, such that
\begin{equation}
    N_\text{hg}= N_\text{lg} (1+V) + N_\text{b} - V (V+1),
\end{equation}
but no interplay between both effects is observed.

\section{Unknown sample orientation \label{ch:rotation}}
\begin{figure}
        \centering
        \includegraphics[width=0.7\linewidth]{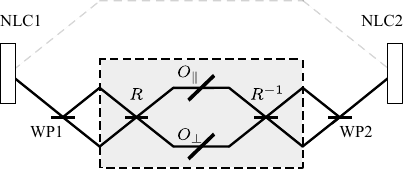}
        \caption{Modified idler arm where the sample is rotated by $\psi$ compared to the initial idler polarization. This rotation can be included via additional SU(2) transformations $R(\psi)$ and $R^{-1}(\psi)$ following and preceding WP1 and WP2 and sandwiching the beam splitter transformations $O_\pp$, $O_\pa$ that account for loss and phase shift of the two perpendicular polarization components of the idler. The gray box indicates the rotated birefringent, diattenuating sample incorporating the rotations by $\psi$ and $-\psi$, respectively, and the object axes. The gray dashed line indicates the signal arm, which remains unaltered.}
        \label{fig:RotatedSample}
\end{figure}

In this section we extend our discussion to probing a sample whose phase-retarding and diattenuating axes are rotated by an unknown angle $\psi$ against the initial idler polarization. 
This situation is particularly relevant for biological sensing since the precise orientation of organic tissue is challenging to control. 
We therefore assume the situation shown in Fig.~\ref{fig:RotatedSample} where a modification of the idler arm is depicted: 
The sample orientation is included by the rotation $R(\psi)$ and its inverse $R^{-1}(\psi)$ before and after the object's axes, respectively.
These rotations can be combined with WP1 and WP2, which directly results in a modification of the transmission and reflection coefficients  $\tau_j$ and $\rho_j$ by the additional parameter $\psi$.
The explicit form of the new transformation coefficients is given in Appendix \ref{ch:sample rotation}.
Since the angle $\psi$ is generally unknown and mixes the two perpendicular axes, $\delta \varphi$ cannot be scanned independently for a rotated sample. 
We thus assume only a scan of the signal arm by tuning the control phase $\phi_0$ as outlined in Sec.~\ref{ch:Fourier} and discuss a fit to obtain the transmission coefficients and phases of a rotated birefringent and diattenuating sample.
To solve for the transmission coefficients independently of the retardance $\delta \varphi$, two scans with different QWP settings are necessary.

Choosing $\gamma_1=\pi/4$, we find that the combination of WP1 with the rotation $R(\psi)$ leads to transformation coefficients $\tau_1(\psi)=\exp(-\i\psi)/\sqrt{2}$ and  $\rho_1(\psi)=\i\exp(\i\psi)/\sqrt{2}$ (see Appendix \ref{ch:sample rotation}).
Hence, this configuration enables the sample rotation to account for an additional phase without introducing further asymmetric attenuation.
This feature arises because WP1 generates circularly polarized light, ensuring that the fraction of light addressing each of the sample's axes is equal.
Choosing WP2 as the inverse of WP1 ($\gamma_2=3\pi/4$), the interferometer output is 
\begin{subequations}
\begin{align}\label{eq:N1}
\begin{split}
    N_\text{lg}^{(1)}=2V\left[ 1 +B_1\sin\left(\bar{\varphi}+\phi_0\right)+ C_1\cos\left(\bar{\varphi}+\phi_0\right)\right],
\end{split}
\end{align} 
where WP2 exactly cancels the dependence on  $\psi$.
A fit to a scan of $\phi_0$ gives rise to the mean phase $\bar \varphi$ and the amplitudes 
\begin{align}
\begin{split}\label{eq:soe1}
  B_1 &= - \frac{\delta t}{2} \sin\frac{\delta \varphi}{2}, \ \ 
  C_1 = \bar{t} \cos\frac{\delta \varphi}{2}.
\end{split}
\end{align}
\end{subequations}
These contain information on the retardance but also the diattenuation of the sample.
Since all three parameters $\delta t$, $\bar t$ and $\delta \varphi$ are independent, they cannot be inferred from the two amplitudes $B_1$ and $C_1$ obtained by fitting.
Therefore, a second setting is needed, where WP1 and WP2 are aligned ($\gamma_1=\pi/4=\gamma_2$) so that we observe the interference signal
\begin{subequations}
\begin{align}\label{eq:N2}
\begin{split}
    N_\text{lg}^{(2)}=2V \left[ 1 +B_2\sin\left(\bar{\varphi}+\phi_0-2\psi\right)+ C_2\cos\left(\bar{\varphi}+\phi_0-2\psi\right) \right]
\end{split}
\end{align}
with the oscillation amplitudes 
\begin{align}
\begin{split}
  B_2 &= -\bar{t}\sin\frac{\delta\varphi}{2},\ \
  C_2 = \frac{\delta t}{2} \cos\frac{\delta \varphi}{2}. \label{eq:soe2}
\end{split}
\end{align}
\end{subequations}
In this configuration the orientation of the sample causes a phase shift of the mean phase.
In contrast to above, not only the parameters $B_2$, $C_2$ but also the sample rotation $\psi$ can be obtained by a fit, where $\bar \varphi$ is known from the previous measurement that led to the interference pattern in form of Eq.~\eqref{eq:N1}.

All transmission and differential phase information can be acquired from Eqs.~\eqref{eq:soe1} and~\eqref{eq:soe2} by solving for the three independent parameters.
Here, the sample rotation $\psi$ and the mean phase $\bar\varphi$ serve as additional fit parameters and we find the relations
\begin{subequations}
\begin{align}
    &\delta \varphi = -2\arctan\left(B_2/C_1\right)\\
    &\bar t = C_1\sqrt{1+(B_2/C_1)^2}, \ \ 
    2\delta t = C_2 \sqrt{1+(B_1/C_2)^2}\label{eq:parameters}
\end{align}
\end{subequations}
for $\delta \varphi \in (-\pi,\pi)$ and $C_1>0$.
For fits leading to $C_1<0$, Eq.~\eqref{eq:parameters} needs to be adjusted by $C_1\rightarrow-C_1$ and $C_2\rightarrow-C_2$ due to $\bar t >0$.
Since $\delta t$ can be negative, $C_2$ can also be negative independently of $\delta \varphi$ and $C_1$.
Thus, the transmission information of an arbitrarily oriented sample exhibiting diattenuation while imprinting a polarization-independent phase can be readily obtained following our proposed WP configurations.

\begin{figure}[t]
        \centering
        \includegraphics[width=0.7\linewidth]{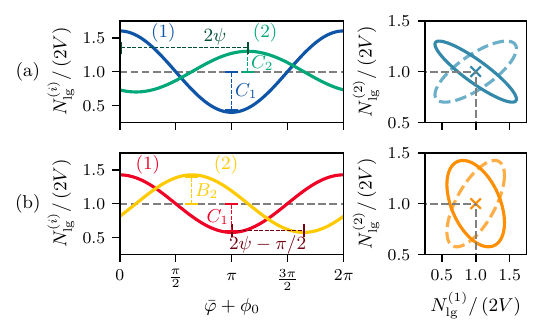}
        \caption{Interference signals $N_{\text{lg}}^{(i)}$ for (a) an isotropically phase-retarding sample with $\delta \varphi = 0$ and (b) polarization-independent attenuation with $\delta t = 0$. We chose in row (a) the values $|t_\s| = 1$, $\delta t = 0.6$, $\bar t = 0.6$ and $\psi = 1.8\,\text{rad}$. In row (b), the settings are $\delta t = 0$ and $\delta \varphi =\pi/2$. We show in the left column the interference signals individually for a scan of the mean phase $\bar \varphi + \phi_0$. The fit parameters $B_2$, $C_1$ and $C_2$, which allow the extraction of $\bar t$, $\delta t$ and $\delta \varphi$, are indicated by the colored dashed lines. The right column depicts the signals $N_{\text{lg}}^{(1)}$ and $N_{\text{lg}}^{(2)}$ in a parametric plot over the control phase $\phi_0$ against each other. The solid ellipses correspond to a sample rotation $\psi = 1.8\,\text{rad}$ while the dashed ellipses correspond to $\psi = 3.5\,\text{rad}$. }
        \label{fig:parametric}
\end{figure}

Sensing an arbitrarily oriented but isotropically phase-retarding sample ($\delta \varphi=0$), shown in Fig.~\ref{fig:parametric}(a), the estimation of the parameters simplifies significantly ($B_1=0=B_2$) and the transmission coefficients can be easily obtained from the fits through
\begin{align}
    \delta t = 2C_2\ \ \text{and} \ \ \bar t = C_1.
\end{align}
Similarly, probing a rotated birefringent sample exhibiting polarization-independent attenuation ($\delta t=0$) \cite{chakraborty2025} as shown in Fig.~\ref{fig:parametric}(b)  we find
\begin{align}\label{eq:fit1}
    \delta \varphi = -2\arctan\left(B_2/C_1\right)\ \ \text{and}\ \ \bar t = C_1\sqrt{1+(B_2/C_1)^2}
\end{align}
for $\delta \varphi \in (-\pi,\pi)$ and $C_1>0$. 
If a fit yields $C_1<0$, $\delta \varphi$ lies outside this interval and is only determined up to a shift of $\pm2\pi$.
Since the mean transmission must be $\bar t>0$, $\bar t$ in Eq.~\eqref{eq:fit1} must then be adjusted by $C_1\rightarrow-C_1$.

In Fig.~\ref{fig:parametric} we show the interference signals Eq.~\eqref{eq:N1} and Eq.~\eqref{eq:N2} for an isotropically phase-retarding sample [row (a)] and for polarization-independent attenuation [row (b)].
The left column depicts the individual signals for a mean phase scan where the corresponding fit parameters are indicated by colored dashed lines.
The transmission and phase information of a rotated birefringent and diattenuating sample can be extracted by performing such fits to experimental data and reading out the parameters $B_i$, $C_j$, $\bar \varphi$ and $\psi$.

The right column depicts parametric plots of $\big(N_\text{lg}^{(1)}, N_\text{lg}^{(2)}\big)^{\text{T}}$ for scans over the control phase $\phi_0$.
As expected for correlated measurements, they result in ellipses centered around $2V(1,1)^{\text{T}}$.
These ellipse shapes also contain the sample information which can be extracted for $\delta \varphi = 0$ and $\delta t = 0$ by performing Lissajous-ellipse fits~\cite{farrell1992, liu2015a, dou2025} to experimental data. 
This process involves rewriting Eqs.~\eqref{eq:N1} and \eqref{eq:N2} in elliptical form, transforming them into a generalized conic section, and fitting the resulting equation to the data.
The conic section's fit parameters then retrieve the sample information.
Such a procedure has the advantage that no knowledge of $\bar \varphi$ is required, so only the sample rotation $\psi$ arises as a fit parameter, since the ellipse's shape and orientation is invariant under global phase shifts. 
The ellipses are, however, highly sensitive to the transmission and rotation properties and phase retardance of the sample.
This is indicated here by the comparison of $\psi =1.8\,\text{rad}$ (solid lines) and $\psi = 3.5\,\text{rad}$ (dashed lines) for both $\delta \varphi =0$ and $\delta t =0$.
Note that ellipse fitting is also applicable for a rotated, birefringent and diattenuating sample by fitting the ellipse $\big(N_\text{lg}^{(1)}, N_\text{lg}^{(2)}\big)^{\text{T}}$ with the five fit parameters $B_i$, $C_i$ and $\psi$ to experimental data.

The presented protocol allows to extract both phase and transmission information of a birefringent and diattenuating rotated sample. 
Since the described fits yield the orientation of the object, the sample can also be aligned with the initial idler polarization by rotating it by $-\psi$ or employing additional retarders, so that the experimental procedures from Sec.~\ref{ch:output} can be applied.

\section{Conclusions}
We theoretically demonstrated the capability of NLIs to simultaneously sense the diattenuation and birefringence properties of samples with undetected light by including waveplates before and after the object.  
These measures are especially relevant for the imaging of biological samples~\cite{mehta2013a, mann2024}, where nonlinear interferometry is particularly advantageous due to its capability of bicolor sensing with undetected photons allowing to address chemically selective frequencies~\cite{diem2013, wrobel2018} while maintaining detection efficiency~\cite{fuenzalida2024a}.

Our proposed scheme ensures that all phase and transmission information is encoded in the polarization of the probing photon which induces coherence. These information are then transferred to the detected mode through nonlinear mixing.
The sample's polarization-dependent phase retardation and light attenuation causes a beating in the resulting three-mode interference signal, and these characteristics of the specimen can be extracted via a Fourier analysis.
We also transferred the operation of the nonlinear processes to the high-gain regime and demonstrated an increase of the visibility in the context of a mean transmission estimation~\cite{michael2021}.
On the other hand, we observed that a phase retardation in the high-gain regime introduced by a birefringent sample causes a complex beating in the output signal due to idler interferences.
Together with the phototoxicity of biological samples at high intensities, this highlights challenges of operating the setup in the strong-pumping regime.
Nevertheless, when the signal photon is blocked between the crystals in this regime, the setup could potentially serve as a nonlinear bicolor analyzer, converting sample information from the interrogating frequency to the detected one without interference in the signal mode, similar to up-conversion imaging~\cite{barh2019, huang2022, junaid2019}.
We also expand our discussion to a sample with unknown orientation by choosing two waveplate configurations which allow Lissajous-ellipse fitting to extract the sample characteristics along with its orientation.

Our results describe the simultaneous sensing of birefringence and diattenuation utilizing NLIs with undetected light while including measurement strategies to extract sample information from experimental data.

\appendix
\section{Transformations \label{ch:transformations}}
In this section we list the transformations used to derive the interferometer output in the signal mode.
All transformation coefficients and operators are specified in Sec.~\ref{ch:setup}.
The action of NLC1 is described by the Bogoliubov transformation of the vacuum input in signal and idler mode
\begin{align}
\hat{b}_\s=u_1\hat{a}_\s+v_1\hat{a}_\i^\dagger    ~\text{ and }~
 \hat{b}_\i=u_1\hat{a}_\i+v_1\hat{a}_\s^\dagger
\end{align}
with $|u_1|^2 - |v_1|^2=1$.
The beam splitter C then phase shifts the signal by the beam splitter transformation
\begin{align}
 \hat{b}'_\s=t_\s \hat{b}_\s+r_\s \hat{l}_\s
\end{align}
with $|t_\s|^2 + |r_\s|^2 =1$.
On the other arm, the idler polarization is  rotated by the first WP.
The subsequent perpendicular and parallel components are obtained from the SU(2) transformation
\begin{align}\label{eq:WP1}
 \hat{c}_\pp=\tau_1\hat{b}_\i+\rho_1\hat{l}_\i
    ~\text{ and }~
 \hat{c}_\pa=-\rho^*_1\hat{b}_\i+\tau^*_1\hat{l}_\i
\end{align}
with $|\tau_1|^2 + |\rho_1|^2 =1$.
These coefficients can be tuned by the settings of the WP, see Appendix~\ref{ch:WP} for their exact form.
The effect of the diattenuating and birefringent object is encoded in additional SU(2) beam splitter transformations, acting individually on the polarization components 
\begin{align}\label{eq:axes}
\hat{c}'_{\pp}=t_\pp \hat{c}_\pp+r_\pp \hat{l}_\pp    ~\text{ and }~
  \hat{c}'_{\pa}=t_\pa \hat{c}_\pa+r_\pa \hat{l}_\pa, 
\end{align}
where the phases $\varphi_\pp=\text{arg}(t_\pp)$ and $\varphi_\pa=\text{arg}(t_\pa)$ are imprinted on the respective polarization.
Again, we have $|t_p|^2 + |r_p|^2 =1$.
The analyzer WP2 rotates the polarization with another SU(2) transformation and only the component 
\begin{align}\label{eq:WP2}
 \hat{b}'_\i=\tau_2\hat{c}'_\pp+\rho_2\hat{c}'_\pa
\end{align}
is used to seed NLC2 and to induce coherence.
Similar to WP1, the coefficients fulfill $|\tau_2|^2 + |\rho_2|^2=1$ and can be tuned according to Appendix~\ref{ch:WP}.
Finally, the modified signal and idler photons seed NLC2, which is again described by a Bogoliubov transformation 
\begin{align}
 \hat{d}_\s=u_2\hat{b}'_\s+v_2\hat{b}'^\dagger_\i
\end{align}
with $|u_2|^2 - |v_2|^2 =1$.
The conventional interferometer phase is given by $\phi_0=\text{arg}(u_1u_2v_1v_2^*t_\s)$.
It contains the pump phase at the two crystals and the phase shift of the signal photon.

 \section{Waveplates as beam splitter transformations \label{ch:WP}}
In this section we describe the action of a waveplate (WP) as an SU(2) transformation. We consider a WP whose fast and slow axes are rotated by the angle $\gamma_i$ compared to the horizontal and vertical axes, which in turn imprint a phase difference $\theta_i$. 
The overall transformation can be written as a sequence of SU(2) transformations and takes the form
\begin{align}
M(\gamma_i,\theta_i)=R^{-1}(\gamma_i)P(\theta_i)R(\gamma_i)
\end{align}
where 
\begin{align}
    R(\gamma_i)=\left(
\begin{array}{cc}
 \cos \gamma_i  & -\sin \gamma_i  \\
 \sin \gamma_i  & \cos \gamma_i  \\
\end{array}
\right)
\end{align} 
describes a rotation by $\gamma_i$ and 
\begin{align}
  P(\theta_i)= \left(
\begin{array}{cc}
 \e^{-\i \frac{\theta_i}{2}} & 0 \\
 0 & \e^{\i \frac{\theta_i}{2}} \\
\end{array}
\right)   
\end{align}
imprints a phase difference $\theta_i$.\\
Introducing the transformation coefficients
\begin{align}\label{eq:r}
 \tau_i=\, \cos^2 \gamma_i \e^{-\i\frac{\theta_i}{2}}+\sin^2 \gamma_i \e^{\i\frac{\theta_i}{2}}    ~\text{ and }~
   \rho_i=\, \i\sin 2\gamma_i \sin\frac{\theta_i}{2},
\end{align}
the transformation can be expressed as 
\begin{equation}
 M(\gamma_i,\theta_i)=\left(
\begin{array}{cc}
 \tau_i & \rho_i \\
 -\rho^*_i & \tau^*_i \\
\end{array}
\right).
\end{equation}
It is easily shown that $|\tau_i|^2+|\rho_i|^2 = 1$ holds.

The phases introduced by two consecutive WPs to the perpendicular $\pp$ and parallel $\pa$ polarization components of the idler photons used in the main body of the article are given by $\phi_\tau=\text{arg}(\tau_{2}\tau_{1})$ and $\phi_\rho=\text{arg}(\rho_{2}\rho_{1}^*)$, respectively.\\

\section{Sample Rotation \label{ch:sample rotation}}
In this section, we incorporate an arbitrary sample orientation into our interferometer model. The effect of a sample rotation by an angle $\psi$ is accounted for using the rotation matrix $R(\psi)$, as illustrated in Fig.~\ref{fig:RotatedSample}.
Hence, we combine WP1 and the rotation $R(\psi)$ before the transformation by the object's axes Eq.~(\ref{eq:axes}) by replacing in our description $M(\gamma_1,\theta_1)\rightarrow R(\psi)M(\gamma_1,\theta_1)$. 
Effectively, this results in the replacements
\begin{subequations}\label{eq:newtransrefl}
\begin{align}
    \begin{split}
    \tau_1 &\rightarrow \tau_1(\psi)=\tau_1 \cos\psi+\rho_1^*\sin\psi\\
    \rho_1 &\rightarrow \rho_1(\psi)=-\tau_1^* \sin\psi+\rho_1\cos\psi
    \end{split}
\end{align}
in the transformations (\ref{eq:WP1}).
Similarly, after the axes we combine WP2 with the reverse transformation  $M(\gamma_2,\theta_2)\rightarrow M(\gamma_2,\theta_2)R^{-1}(\psi)$ resulting in the replacements
\begin{align}
    \begin{split}
    \tau_2 &\rightarrow \tau_2(\psi)=\tau_2 \cos\psi-\rho_2\sin\psi\\
    \rho_2 &\rightarrow \rho_2(\psi)=\tau_2 \sin\psi+\rho_2\cos\psi
    \end{split}
\end{align}
\end{subequations}
in transformation (\ref{eq:WP2}). The remaining transformations are invariant under a sample rotation.

\begin{backmatter}
\bmsection{Acknowledgment}
We thank Jorge Fuenzalida, Markus Gräfe, Jonas Moos, Sebastian Töpfer and Sergio Adrián Tovar Pérez for fruitful discussions.

\bmsection{Disclosures}
The authors declare no conflicts of interest.

\bmsection{Data availability} No data were generated or analyzed in the presented research.
\end{backmatter}
\bibliography{references}

\end{document}